\documentclass[fleqn,12pt,twoside]{article}
\usepackage{espcrc1}
\input BoxedEPS
\SetRokickiEPSFSpecial  
\HideDisplacementBoxes

\def\rd#1{\mathop{{\rm d}#1}}
\def\braket#1#2#3{\bigl\langle #1 \bigl| #2 \bigr| #3\bigr\rangle}

\newcommand{\AmS}{{\protect\the\textfont2
   A\kern-.1667em\lower.5ex\hbox{M}\kern-.125emS}}

\title{Understanding Parton Distributions from Lattice QCD:\\
Present Limitations and Future Promise
\thanks{Work supported in
part by the U.S. Department of Energy (DOE) under cooperative 
research agreement
\#DE-FC02-94ER40818. \quad MIT-CTP-3162}}

\author{ J. W. Negele\address
{Center for Theoretical Physics,
Massachusetts Institute of Technology, \\
         77 Massachusetts Avenue, Cambridge, Massachusetts 02139, USA}}

\begin{document}

\maketitle

\begin{abstract}

This talk will explain how ground state matrix elements specifying 
moments of quark
density and spin distributions in the nucleon  have been calculated 
in full QCD,  show how
physical extrapolation to the chiral limit including  the physics of 
the pion cloud
resolves previous apparent conflicts with experiment, and
describe the computational resources required for a
definitive comparison with experiment.

\end{abstract}

\section{INTRODUCTION}

The focus of this workshop, the QCD structure of the nucleon, has two 
complementary aspects.
One is measuring the quark and gluon structure experimentally. 
Decades of impressive high
energy scattering experiments utilizing our knowledge of perturbative 
QCD and factorization have
produced detailed  knowledge of the light cone
distributions of quarks and gluons in the nucleon, with the promise 
of even richer
phenomenology to come. The second, and presently less well developed 
aspect, is calculating and
understanding this quark and gluon structure theoretically, and the 
only method known at
present to solve nonperturbative QCD is lattice field theory. Both 
aspects are essential, and I will
show that the tools for quantitative calculation of nucleon structure 
are at hand, explain the
scale of computational resources required for a definitive 
calculation, and argue that our field
needs to provide these computational resources as well as 
experimental resources if we are to
truly understand hadron structure.

Using the
operator product expansion and lattice field theory, it is possible 
to calculate
moments of quark distributions, and I will discuss here the first
calculations in full QCD\cite{dolgov-thesis,Dolgov:2002zm}. A major 
puzzle in the field
has been the fact that quenched calculations of these moments, which ignore
quark-antiquark excitations of the Dirac sea, disagree with experiment at the
20-50\% level. I will show that contrary to earlier conjectures, at 
the quark masses
accessible in practical calculations, including quark loops does not 
alter the results
significantly. Rather,  I will  argue that the physical origin of the 
discrepancy with
experiment has been incorrect extrapolation to the physical quark 
mass, and will show
how extrapolation incorporating the leading non-analytic behavior 
required by chiral
symmetry produces consistent results for the moments of quark distributions. In
addition, we have  also  compared full QCD results with 
configurations that have been
cooled to remove all the gluon contributions except for those of 
instantons and shown
that the qualitative behavior of the moments is reproduced by the 
instanton content
of the gluon configurations.

\section{MOMENTS OF QUARK DISTRIBUTIONS IN THE PROTON}

\begin{figure}[tp]
\vspace*{-2mm}
$$
\BoxedEPSF{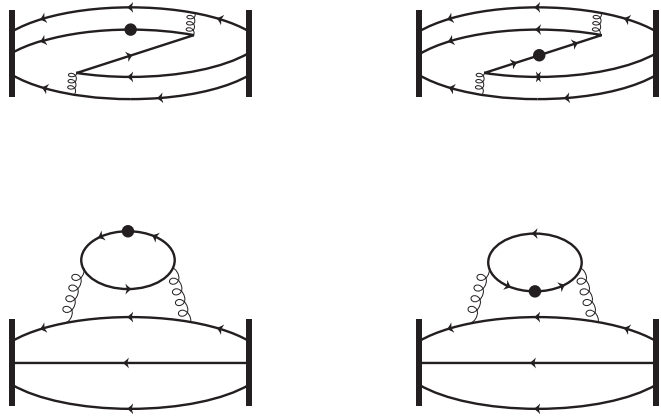 scaled 900}
$$
\vspace*{-.6cm }
\caption{Connected (upper row) and disconnected (lower row) diagrams 
contributing
to hadron matrix elements. The left column shows typical 
contributions of quarks and
the right column shows contributions of antiquarks}
\label{HadrMatrElem}
\vspace*{-.5cm}
\end{figure}

By the operator product expansion, moments of the following linear 
combinations of
quark and antiquark distributions in the proton
\begin{eqnarray}
\langle x^n\rangle_q &=&
\int_0^1 \rd x x^n \bigl(q (x) + (-1)^{n+1}\bar q(x)\bigr)  \label{mom1}\\
\langle x^n\rangle_{\Delta q} &=&
\int_0^1 \rd x x^n \bigl(\Delta q (x) + (-1)^{n}\Delta\bar q(x)\bigr) 
\nonumber\\
\langle x^n\rangle_{\delta q} &=&
\int_0^1 \rd x x^n \bigl(\delta q (x) + (-1)^{n+1}\delta\bar 
q(x)\bigr) \nonumber
  \end{eqnarray}
are related to the following matrix elements of twist-2 operators
\begin{eqnarray}
\braket {PS}{\bar\psi \gamma^{\{\mu_1} i  D^{\mu_2} \cdots i
D^{\mu_n\}}
\psi}{PS} &=& 2 \langle x^{n-1}\rangle_q\, P^{\{\mu_1}\cdots P^{\mu_n\}}
\label{mom2}\\
\braket {PS}{\bar\psi \gamma^{\{\mu_1}\gamma_5 i  D^{\mu_2} \cdots i
D^{\mu_n\}}
\psi}{PS} &=& 2 \langle x^{n-1}\rangle_{\Delta q}\,
MS^{\{\mu_1}P^{\mu_2}\cdots P^{\mu_n\}} \nonumber\\
\braket{PS}{\bar\psi \sigma^{[\alpha\{\mu_1]}\gamma_5 i
D^{\mu_2}
\cdots i D^{\mu_n\}}
\psi}{PS} &=& 2 \langle x^{n-1}\rangle_{\delta q}\,
MS^{[\alpha}P^{\{\mu_1]}P^{\mu_2}\cdots P^{\mu_n\}} \nonumber.
\end{eqnarray}
Here,   $q = q_\uparrow + q_\downarrow ,  \Delta q =
q_\uparrow - q_\downarrow$,  $\delta q = q_\top + q_\bot $, $x$ denotes the
momentum fraction carried by the quark, and  $\{\,\} $ and $[\,] $ denote
symmetrization and  antisymmetrization respectively. We note that odd moments
$\langle x^n\rangle_q$ are obtained from deep inelastic electron or 
muon scattering
structure functions
$F_1$ or
$F_2$,  even moments of $\langle x^n\rangle_{\Delta q} $ are determined from
$g_1$, and these moments are proportional to the quantities $v_{n+1}$ and $a_n$
defined in ref.\cite{qcdsf}. In addition, $g_2$ also determines the 
quantity $d_n$
\begin{equation}
\langle PS |  \bar{\psi}
\gamma^{[\sigma}\gamma_5iD^{\{\mu_1]} \cdots
iD^{\mu_n\}}\psi | PS\rangle = \frac{1}{n} {d_n
S^{[\sigma}P^{\{\mu_1]}\cdots P^{\mu_n\}}}
\end{equation}
although the lattice calculation of $d_n$ with Wilson fermions is 
complicated by mixing with the
lower dimension operator $ \frac{1}{a}\gamma^{[\sigma}\gamma_5 
\gamma^{\{\mu_1]} \cdots
iD^{\mu_n\}}$.

Even moments  $\langle
x^n\rangle_q$ are obtained from deep inelastic neutrino scattering, 
and in addition, a
variety of other processes have contributed to what is now a detailed empirical
knowledge of the quark and antiquark distributions in the nucleon. We will
subsequently compare our  results with moments calculated from the 
CTEQ, MRS, and
GRV global fits to the world supply of data.

\section{CALCULATION OF MATRIX ELEMENTS}

Proton matrix elements of the operators in Eq.\ref{mom2} are calculated by
evaluating the connected and disconnected diagrams shown in Fig.
\ref{HadrMatrElem}. Note that both the connected and disconnected diagrams each
receive contributions from quarks and antiquarks. Depending on the moment, by
Eq.\ref{mom1}, the sum of the diagrams yields either the sum or 
difference of the
moments of the quark and antiquark distributions.  Because it is 
technically much more
difficult to evaluate the disconnected diagrams, our present 
calculations only include
connected diagrams. Fortunately, the disconnected diagrams are flavor 
independent,
so they cancel out of the flavor non-singlet difference between up 
and down quark distributions.
Hence, in Table
\ref{tab-summary},  we compare lattice calculation of the difference
between connected diagrams for up and down quarks with the corresponding
difference  of moments of experimental data for the sum or difference 
of quark and
antiquark distributions.
%
%

%

\begin{figure}[tp]
\vspace*{-2mm}
$$
\BoxedEPSF{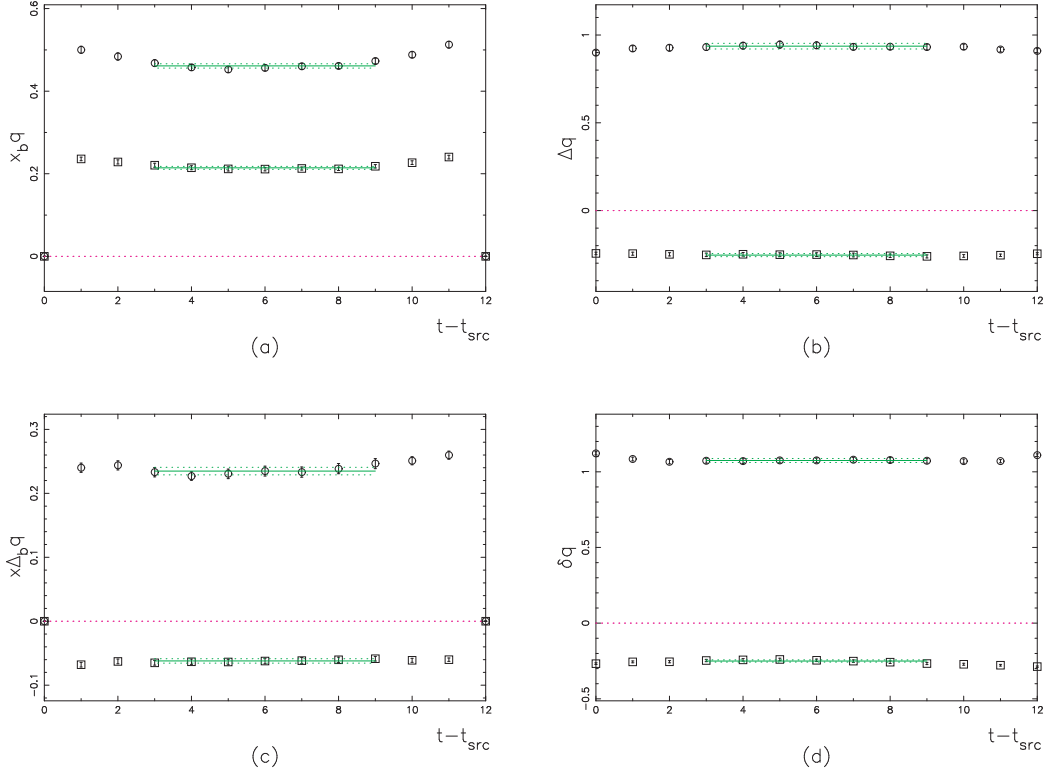 scaled 550}
$$
\vspace*{-2pc}
\caption{Plateaus in measurements of operators in a zero momentum 
ground state as
a function of the euclidean time separation from the source.}
\label{NewFigs1}
\vspace*{-.75cm}
\end{figure}

On the lattice, connected diagrams are evaluated by calculating a 
three point function
in which a source creates a state with the quantum numbers of the proton,
the operator acts on this state, and a sink finally annihilates the state.
Because evolution in imaginary time filters out the ground state,  when the
operator is sufficiently far from both the source and sink, it acts in the
ground state and produces the desired ground state matrix element. As 
the time at
which it acts approaches either the source or sink, it sees excited 
state contaminants,
yielding a central plateau corresponding to the physical matrix element and
exponential contaminants at the edges. Obviously, it is beneficial to 
optimize the
overlap of the source with the ground state to maximize the plateau region and
minimize the effect of the excited state contaminants at the edges.

In this work, connected diagrams were calculated using sequential
propagators generated by the upper two components of the
nucleon source $J^{\alpha} =
u_a^{\alpha}u_b^{\beta}(C\gamma_5)_{\beta,\beta'}
d_c^{\beta'}\epsilon^{abc}$. The overlap with the physical
proton ground state was optimized using Wuppertal
smearing \cite{sources} to maximize the  overlap  $P(0) = {|\langle 
J|0\rangle |^2}$.
Varying the smearing reduced $P$ by over 4
orders of magnitude, yielding an overlap with the physical
ground state of approximately 50\%. Dirichlet boundary
conditions were used for quarks in the t-direction.

The resulting plateaus for four operators that could be measured in a 
proton with
zero three-momentum are shown in Fig.~\ref{NewFigs1}. Here one observes both a
statistically well determined central plateau region and the 
effectiveness with which
the excited state contaminants have been reduced by the optimized source.
\vspace*{-\medskipamount}

\section{OPERATORS AND
PERTURBATIVE RENORMALIZATION}

\begin{table}[tp]
\caption{Lattice Operators}
\begin{tabular}{llrl}
\hline
& H(4) mix&$\vec{p}$&lattice operator\\
\hline
$\langle x \rangle _q^{(a)}$ & {\bf 6}$_3^+$ \, no & $\neq 0$  & 
$\bar{q} \gamma_{\{1}
{\stackrel{\,\leftrightarrow}{D}}_{4\}} {q}$  \\
$\langle x \rangle _q^{(b)}$ & {\bf 3}$_1^+$ \, no & 0& $\bar{q} \gamma_{4}
{\stackrel{\,\leftrightarrow}{D}}_{4} {q}- \frac{1}{3} \sum_{i=1}^3
  \bar{q} \gamma_{i} {\stackrel{\,\leftrightarrow}{D}}_{i} {q}$ \\
$\langle x ^2 \rangle _q$     & {\bf 8}$_1^-$ \, yes & $\neq 0$ & 
$\bar{q} \gamma_{\{1}
{\stackrel{\,\leftrightarrow}{D}}_{1}
{\stackrel{\,\leftrightarrow}{D}}_{4\}} {q} - \frac{1}{2} \sum_{i=2}^3
\bar{q}\gamma_{\{i} {\stackrel{\,\leftrightarrow}{D}}_{i}
{\stackrel{\,\leftrightarrow}{D}}_{4\}}{q}$ \\
$\langle x^3 \rangle _q$    & {\bf 2}$_1^+$ \, no$^*$ & $\neq 0$ & 
$\bar{q} \gamma_{\{1}
{\stackrel{\,\leftrightarrow}{D}}_{1}
{\stackrel{\,\leftrightarrow}{D}}_{4}
{\stackrel{\,\leftrightarrow}{D}}_{4\}} {q} + \bar{q} \gamma_{\{2} 
{\stackrel{\,\leftrightarrow}{D}}_{2}
{\stackrel{\,\leftrightarrow}{D}}_{3}
{\stackrel{\,\leftrightarrow}{D}}_{3\}} {q} - (\,3\,\leftrightarrow\,4\,)$ \\
\hline
$\langle 1 \rangle _{\Delta q}$     & {\bf 4}$_4^+$ \, no & 0  & $\bar{q}
\gamma^{5} \gamma_{3} {q}$  \\
$\langle x \rangle _{\Delta q}^{(a)}$  & {\bf 6}$_3^-$  \, no & $\neq 
0$ & $\bar{q}
\gamma^5 \gamma_{\{1} {\stackrel{\,\leftrightarrow}{D}}_{3\}} {q}$ \\
$\langle x \rangle _{\Delta q}^{(b)}$ & {\bf 6}$_3^-$  \, no & 0 & 
$\bar{q} \gamma^5
\gamma_{\{3} {\stackrel{\,\leftrightarrow}{D}}_{4\}} {q}$ \\
$\langle x^2 \rangle _{\Delta q}$   & {\bf 4}$_2^+$  \, no & $\neq 0$ 
& $\bar{q}
\gamma^5 \gamma_{\{1} {\stackrel{\,\leftrightarrow}{D}}_{3}
{\stackrel{\,\leftrightarrow}{D}}_{4\}} {q}$ \\
\hline
$\langle 1 \rangle _{\delta q}$      & {\bf 6}$_1^+$  \, no & 0  & 
$\bar{q} \gamma^{5}
\sigma_{34}  {q}$  \\
$\langle x \rangle _{\delta q}$   & {\bf 8}$_1^-$ \, no & $\neq 0$ 
& $\bar{q} \gamma^{5}
\sigma_{3\{4} {\stackrel{\,\leftrightarrow}{D}}_{1\}}{q}$  \\
$d_1$     & {\bf 6}$_1^+$  \, no$^{**}$ & 0  & $\bar{q} \gamma^5 \gamma_{[3}
{\stackrel{\,\leftrightarrow}{D}}_{4]}{q}$ \\
$d_2$     & {\bf 8}$_1^-$  \, no$^{**}$ & $\neq 0$ & $\bar{q} \gamma^5
\gamma_{[1} {\stackrel{\,\leftrightarrow}{D}}_{\{3]}
{\stackrel{\,\leftrightarrow}{D}}_{4\}} {q}$ \\
\hline
\end{tabular}
\end{table}
\begin{table}[hbt]
\caption{Perturbative renormalization}  
\begin{tabular}{lcrcll}
\hline\\[-4.5mm]
& $\gamma$ & $B^{LATT}$ & $B^{{\overline{MS}}}$ & $Z_{\beta=6.0}$& 
 $Z_{\beta=5.6}$\\ 
\hline
$\langle x \rangle _q^{(a)}$         &     $\frac{8}{3}$  & 
$-3.16486$  &      $-\frac{40}{9}$
    & $0.989$& $0.988$ \\
$\langle x \rangle _q^{(b)}$         &     $\frac{8}{3}$  & 
$-1.88259$  &      $-\frac{40}{9}$
    & $0.978$& $0.977$ \\
$\langle x ^2 \rangle _q$            &    $\frac{25}{6}$  & 
$-19.57184$  &      $-\frac{67}{9}$
    & $1.102$& $1.110$ \\
$\langle x^3 \rangle _q$            &  $\frac{157}{30}$  & 
$-35.35192$  &  $-\frac{2216}{225}$
    & $1.215$& $1.231$ \\
$\langle 1 \rangle _{\Delta q}$        &               $0$  & 
$15.79628$  &                  $0$
    & $0.867$& $0.857$ \\
$\langle x \rangle _{\Delta q}^{(a)}$ &     $\frac{8}{3}$  & 
$-4.09933$  &      $-\frac{40}{9}$
    & $0.997$& $0.997$ \\
$\langle x \rangle _{\Delta q}^{(b)}$  &     $\frac{8}{3}$  & 
$-4.09933$  &      $-\frac{40}{9}$
    & $0.997$& $0.997$ \\
$\langle x^2 \rangle _{\Delta q}$      &    $\frac{25}{6}$  & 
$-19.56159$  &      $-\frac{67}{9}$
    & $1.102$& $1.110$ \\
$\langle 1 \rangle _{\delta q}$       &               $1$  & 
$16.01808$  &                 $-1$
    & $0.856$& $0.846$ \\
$\langle x \rangle _{\delta q}$        &               $3$  & 
$-4.47754$  &                 $-5$
    & $0.996$& $0.995$ \\
$d_1$              &               $0$  &  $0.36500$  &                  $0$
    & $0.997$& $0.997$ \\
$d_2$              &     $\frac{7}{6}$  &  $-15.67745$  &     $-\frac{35}{18}$
    & $1.116$& $1.124$ \\
\hline
\end{tabular}
\end{table}

\vspace*{.5cm}

The continuum operators defined above are approximated on a
discrete cartesian lattice using representations of the
hypercubic group that have been chosen to eliminate operator mixing as much as
possible and to  minimize the number of non-zero
components of the
nucleon momentum.  The
operators we have used are shown in Table~1, 
where we
have indicated whether the spatial momentum components
are non-zero and whether mixing occurs. Note, no$^*$
indicates a case in which mixing could exist in general but
vanishes perturbatively for Wilson or overlap fermions and
no$^{**}$ indicates perturbative mixing with {\it lower}
dimension operators for Wilson fermions but no mixing for
overlap fermions.  Because the statistical errors are much larger for sources
projected to non-zero momentum, the moments corresponding to 
operators requiring
non-zero momentum are presently not well determined.

To convert from  lattice regularization at the scale of the inverse 
lattice spacing $1/a$
to the continuum $\overline{MS}$ scheme at momentum scale Q, we use 
the one-loop
perturbation theory result
\begin{displaymath}
O^{\overline{MS}}_i(Q^2)=\sum_j\left(\delta_{ij}+\frac{g_0^2}{16\pi^2}\,
\frac{N_c^2-1}{2N_c}
\left(\gamma^{\overline{MS}}_{ij}\log(Q^2a^2)-(B^{LATT}_{ij}
-B^{\overline{MS}}_{ij})\right)\right)\cdot O^{LATT}_j(a^2) .
\label{app-renorm1}
\end{displaymath}
The anomalous dimensions $\gamma_{ij}$ and the finite constants $B_{ij}$
we have
calculated and used in this work are tabulated in
Table~2 \cite{pert-renorm}.
\bigskip

\begin{figure}[t]
\vspace*{-8mm}
$$
\BoxedEPSF{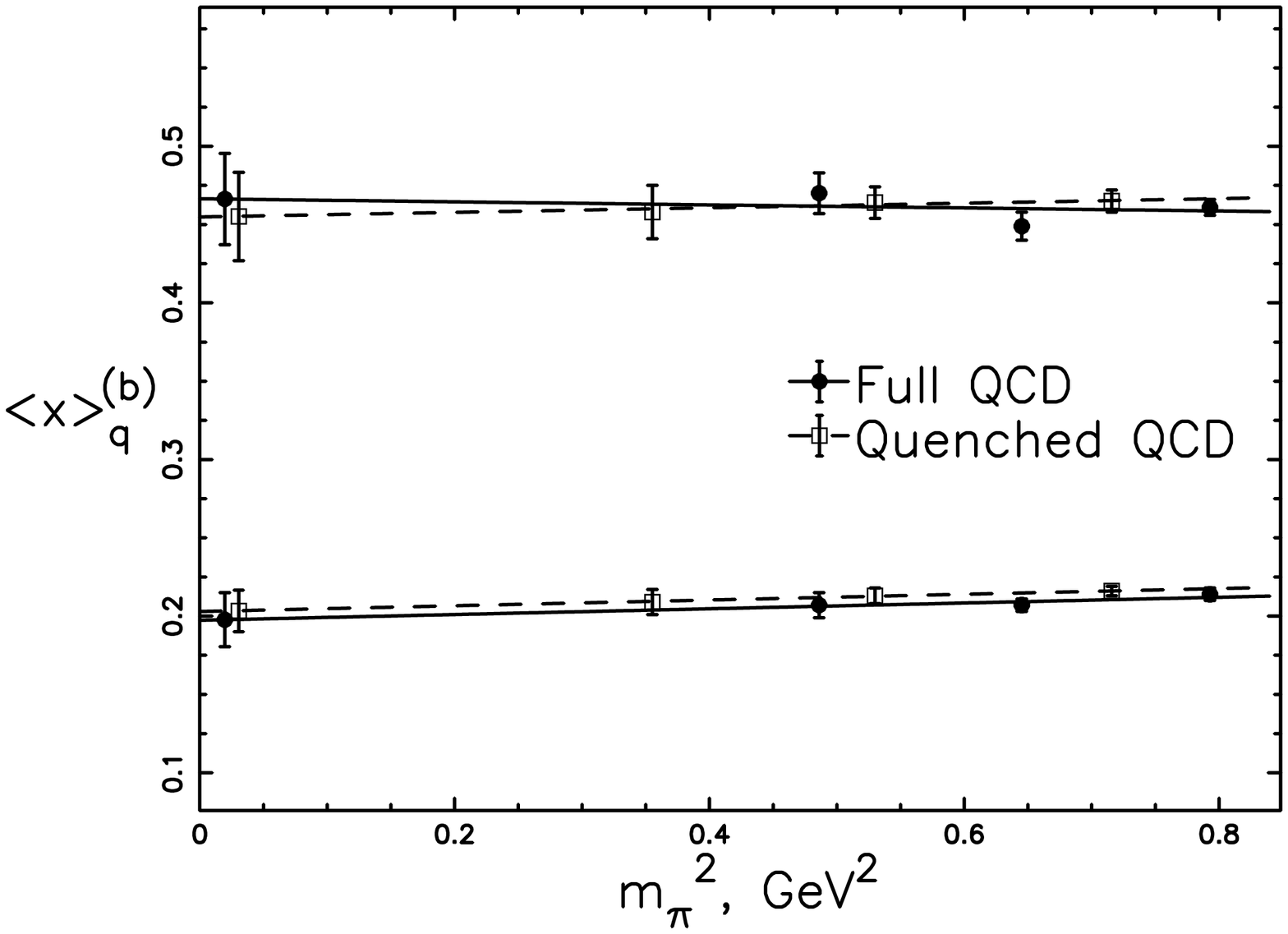 scaled 400}
\quad
\raise4pt \hbox{\BoxedEPSF{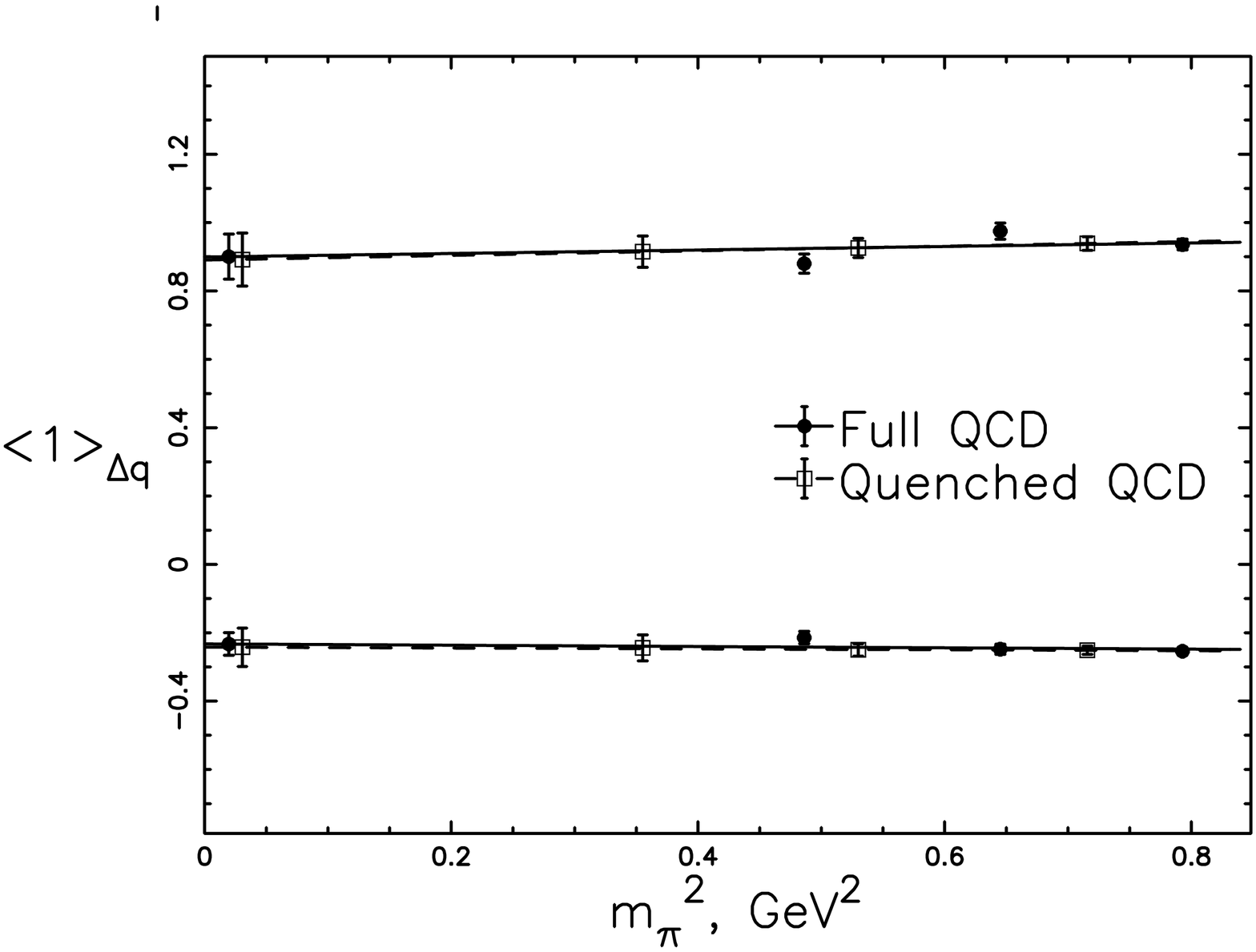 scaled 370} }
$$
\vspace*{-2pc}
\caption{Comparison of linear chiral extrapolations of full and quenched
calculations of $\langle x \rangle _q^{(b)}$  and  $\langle 1 \rangle 
_{\Delta q}$
showing
agreement within statistical errors  }
  \label{fig-full}
\vspace*{-.5cm}
\end{figure}

\begin{figure}[b]
\vspace*{-8mm}
$$
\BoxedEPSF{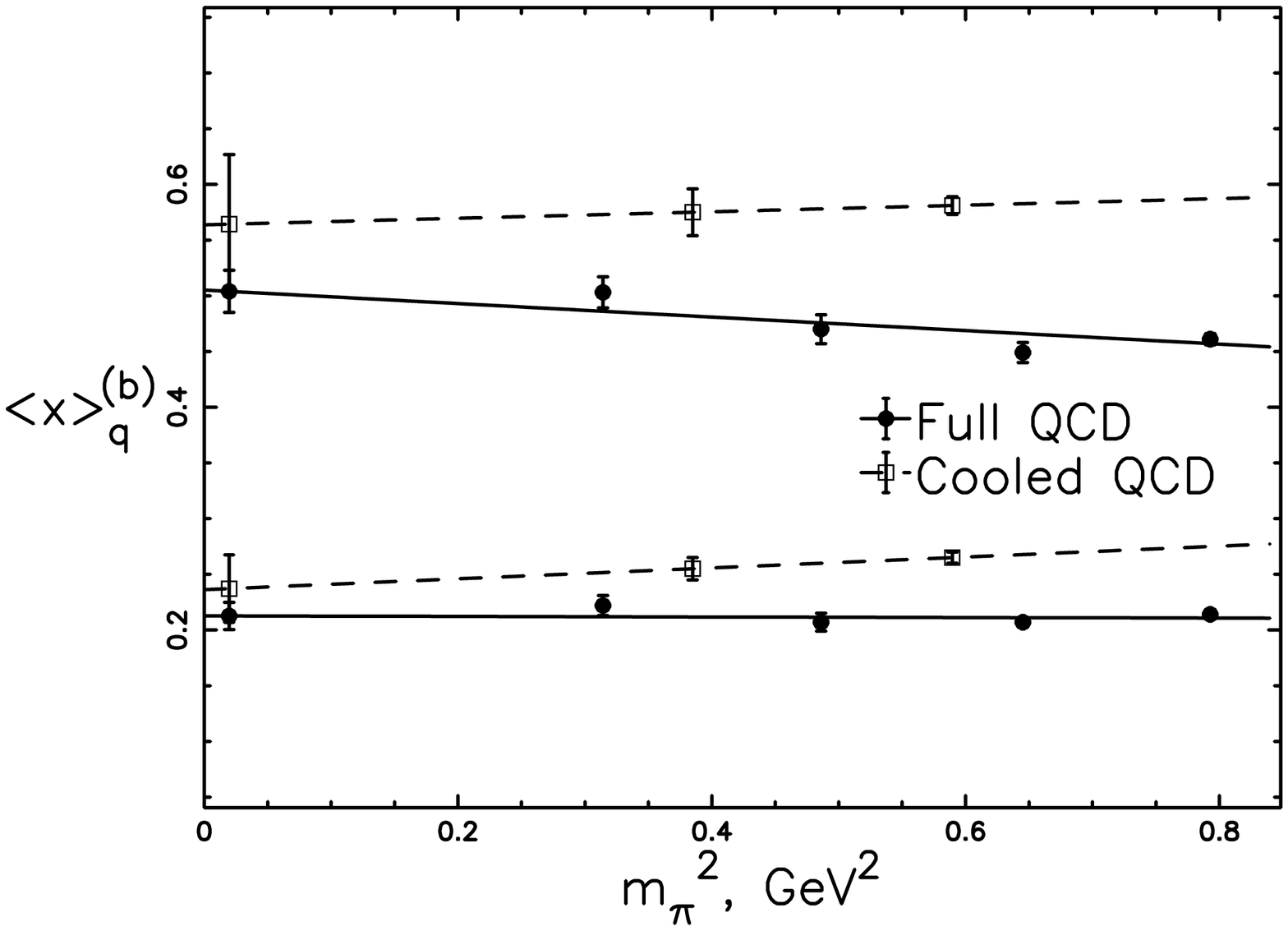 scaled 400}
\quad
\BoxedEPSF{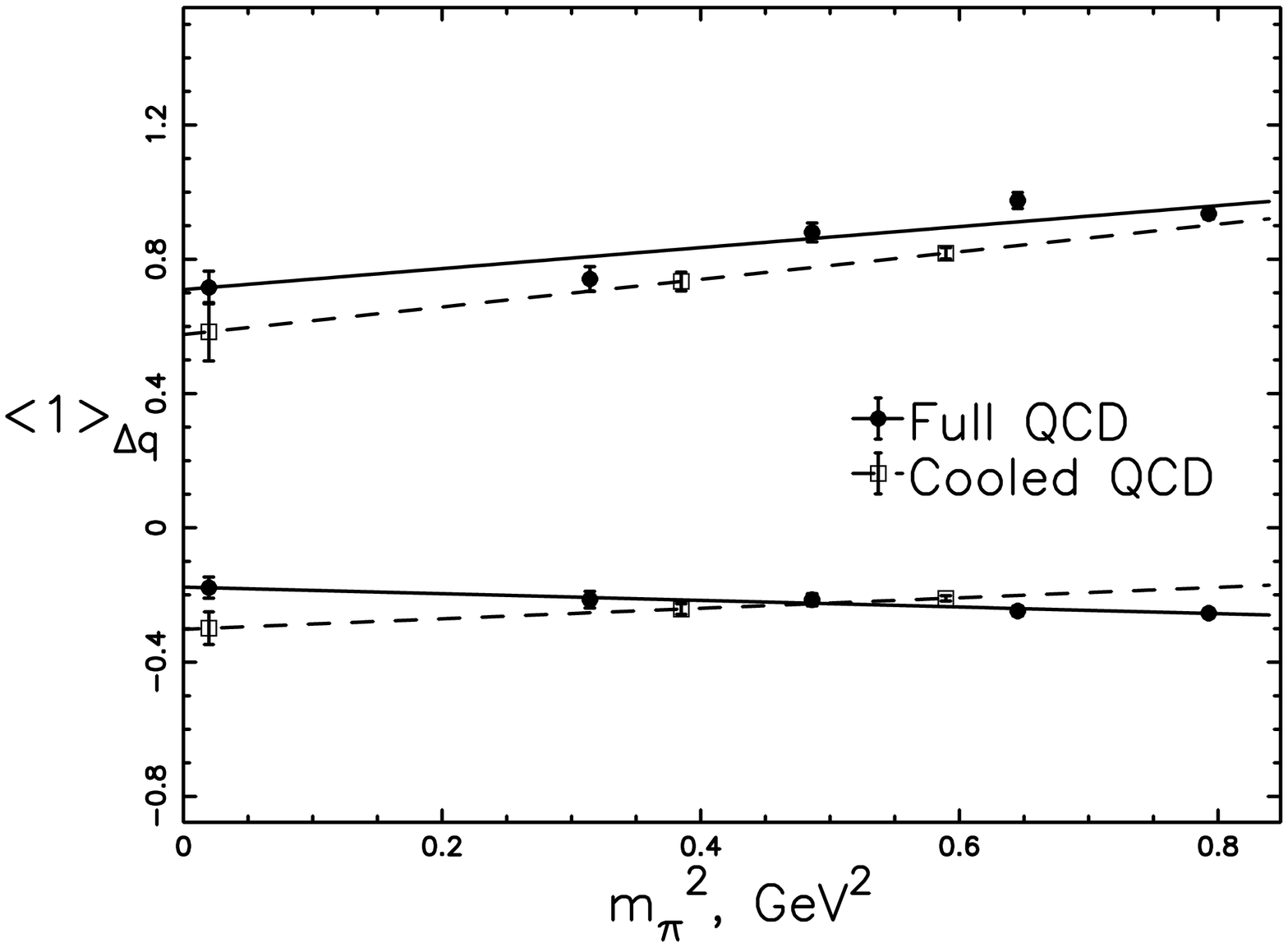 scaled 390}
$$
\vspace*{-2pc}
\caption{Comparison of linear chiral extrapolations of full and cooled
calculations of $\langle x \rangle _q^{(b)}$  and  $\langle 1 \rangle 
_{\Delta q}$ showing the
extent to which instantons reproduce the full result.}
\label{fig-full-cooled}
\vspace*{-.5cm}
\end{figure}

\begin{table}[hbt]
\caption{Comparison of linear extrapolations of our full QCD and 
quenched results with
other lattice calculations and phenomenology at  $4$ GeV in $\overline{MS}$}
\label{tab-summary}
\begin{center}
\begin{tabular}{lrrrrrr}
\hline
Conn. & QCDSF & QCDSF & SESAM &Quenched &Full QCD
& \multicolumn{1}{c}{Phen.} \\
  M. E.&  & ($a=0$) &  &  & (3 pts) & \multicolumn{1}{c}{($q\pm \bar q$)} \\
\hline
$ \langle x \rangle_u $      & $0.452(26)$ &   &   & $0.454(29)$ & 
$0.459(29)$ &  \\
$\langle x \rangle_d $      & $0.189(12)$ &   &   & $0.203(14)$ & 
$0.190(17)$ &  \\
$\langle x \rangle_{u-d}$    & $0.263(17)$ &   &   & $0.251(18)$ & 
$0.269(23)$ &
$0.154 (3)$ \\
$\langle x^2 \rangle_u $    & $0.104(20)$ &   &   & $0.119(61)$ & 
$0.176(63)$ &  \\
$ \langle x^2 \rangle_d $    & $0.037(10)$ &   &   & $0.029(32)$ & 
$0.031(30)$ &
\\
$\langle x^2 \rangle_{u-d} $ & $0.067(22)$& &  & $0.090(68)$ &  $0.145(69) $
& $0.055 (1) $\\
$ \langle x^3 \rangle_u $    & $0.022(11)$ &   &   & $0.037(36)$ & 
$0.069(39)$ &
  \\
$ \langle x^3 \rangle_d $    & $-0.001(7)$ &   &   & $0.009(18)$ & $-0.010(15)$
  \\
$\langle x^3 \rangle_{u-d} $ & $0.023(13)$& &  & $0.028(49)$ &  $0.078(41) $
& $0.023 (1) $\\
$ \langle 1\rangle_{\Delta u}$ & $0.830(70)$ & $0.889(29)$ & $0.816(20)$
& $0.888(80)$ & $0.860(69)$ &  \\
$\langle 1\rangle_{\Delta d}$ & $-0.244(22)$ & $-0.236(27)$ & $-0.237(9)$
& $-0.241(58)$ & $-0.171(43)$ &  \\
$ \langle 1\rangle_{\Delta u -\Delta d} $ & $1.074(90)$ & $1.14(3)$ & 
$1.053(27)$
& $1.129(98)$ & $1.031(81)$ & $1.248 (2)$ \\
$\langle x \rangle_{\Delta u} $ & $0.198(8)$ &   &  & $0.215(25)$ & 
$0.242(22)$ & \\
$\langle x \rangle_{\Delta d}$ & $-0.048(3)$ &   &  & $-0.054(16)$ &
$-0.029(13)$ & \\
$\langle x \rangle_{\Delta u -\Delta d}$ & $0.246(9)$ &      &  & $0.269(29)$ &
$0.271(25)$ & $0.196 (9)$ \\
$ \langle x ^2\rangle_{\Delta u} $ & $0.04(2)$ &      &  & $0.027(60)$ &
$0.116(42)$ & \\
$ \langle x ^2\rangle_{\Delta d} $ & $-0.012(6)$ &      &  & $-0.003(25)$ &
$0.001(25)$ &\\
$ \langle x ^2\rangle_{\Delta u - \Delta d} $ & $0.05(2)$ &      &  & 
$0.030(65)$
&
$0.115(49)$ & $0.061 (6)$ \\
$ \delta u_c $ & $0.93(3)$ & $0.980(30)$ &   & $1.01(8)$ & $0.963(59)$ &  \\
$ \delta d_c $ & $-0.20(2)$ & $-0.234(17)$ &   & $-0.20(5)$ & 
$-0.202(36)$ &  \\
$ d_2^u $ & $-0.206(18)$ &     &   & $-0.233(86)$ & $-0.228(81)$ &  \\
$ d_2^d $ & $-0.035(6)$ &     &   & $0.040(31)$ & $0.077(31)$ &  \\
\hline
\end{tabular}
\end{center}
\vspace*{-3pc}
\end{table}

\section{RESULTS}

The moments listed in Table 1 were calculated 
\cite{dolgov-thesis,Dolgov:2002zm}
on $16^3 \times 32$ lattices for Wilson
fermions in full QCD at $\beta = 5.6$ using 200 SESAM configurations
at each of 4 $\kappa 's$ and at $\beta = 5.5$ using 100 SCRI
configurations at 3 $\kappa 's$.
They were also calculated with two sets of 100 full
QCD configurations cooled with 50 cooling steps and in
quenched  QCD at $\beta = 6.0$ using 200 configurations at each of 3
$\kappa 's$.  Typical linear chiral extrapolations for operators calculated
with nucleon momentum equal to zero are shown in Figure~\ref{fig-full}
for full and quenched
calculations of $\langle x \rangle_q$ and  $\langle x
\rangle_{\Delta q}$, showing agreement within statistical errors. To avoid
finite volume errors at the
lightest quark mass, the SESAM \cite{sesam} results were extrapolated using the
three heaviest quark masses. Table~3 shows a major result of our work,
that there is complete
agreement within statistics between full and quenched
results. Statistics with the SCRI configurations \cite{scri}
are currently being increased to study the dependence on the coupling constant.

Typical
chiral extrapolations for cooled configurations are compared with the
corresponding uncooled full QCD calculations in
Figure~\ref{fig-full-cooled}.
This qualitative agreement between cooled and uncooled results occurs at
light quark mass for
all the twist-2 matrix elements we calculated  and demonstrates the
degree to which the instanton content of the configurations
and their associated zero modes dominate light hadron structure
\cite{instantons}.

\begin{figure}[t]
\vspace*{-2mm}
$$
\BoxedEPSF{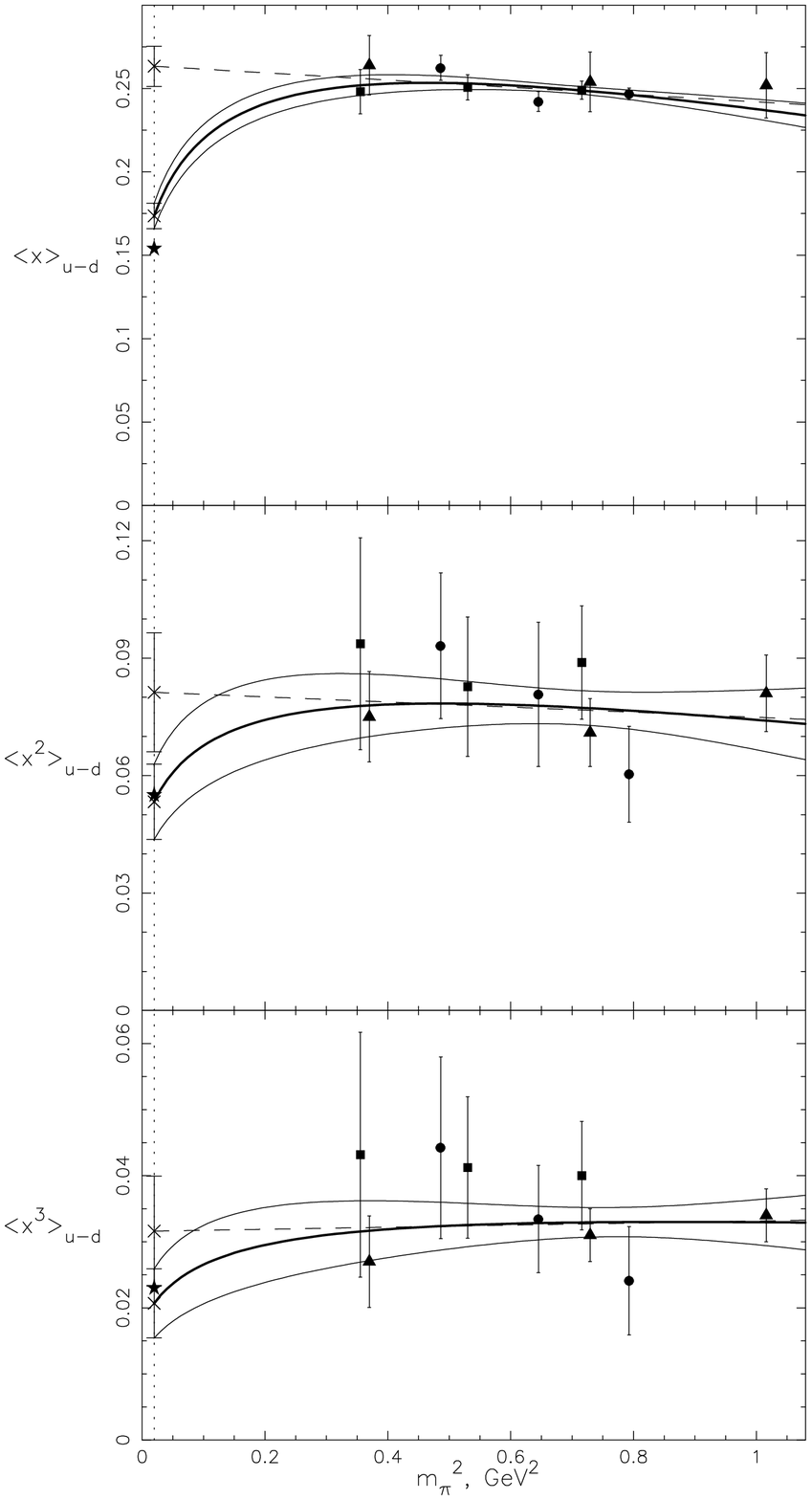 scaled 500}
$$
\vspace*{-.5cm}
\caption{Chiral extrapolation of $\langle x \rangle _q^{(b)}$ , 
$\langle x ^2 \rangle _q$ , and
$\langle x^3 \rangle _q$
using Eq. \ref{chiral_ext}}
\label{fit}
\end{figure}

A major crisis and longstanding puzzle in this field has been the 
fact that when quenched lattice
results are linearly extrapolated in $m_q$,  results disagree with 
experiment at the 20-50\%
level. Our results clearly show that inclusion of quark-antiquark 
excitation from the Dirac Sea
does not resolve this discrepancy.  Salient examples from Table 
\ref{tab-summary}	are
$\langle x \rangle_{u-d} \sim 0.25 - 0.27$  compared with the 
experimental result
0.15 and $ g_A =
\langle 1\rangle_{\Delta u -\Delta d}  \sim 1.0 - 1.1$  compared with 
the experimental
result 1.26.

A breakthrough in resolving this crisis has been the understanding that
the physical origin of these discrepancies is the
inadequate treatment of the pion cloud in the nucleon that has been necessary
because of limited computational resources. By necessity, present 
calculations are
restricted to quark masses that are so heavy that the pion mass is 
above 600 MeV
and the pion cloud surrounding the nucleon is strongly suppressed. Physical
quantities like the nucleon magnetic moment and axial charge clearly 
depend strongly
on the pion current, and should therefore be very sensitive to the 
absence of the full
pion cloud. Furthermore, because of the rapid, nonlinear variation 
from the chiral logs
arising from Goldstone boson loops, it is clear that a linear 
extrapolation is completely
inadequate to describe the correct chiral physics.

  In a recent work\cite{Detmold:2001jb},  we have shown that chiral 
extrapolation incorporating
the leading non-analytic behavior from chiral perturbation theory can 
systematically resolve the
discrepancies in the moments $\langle x^n \rangle_{u-d}$ using the formula:
\begin{equation}
\langle x^n  \rangle_u - \langle x^n  \rangle_d \sim a_n \left[ 1 - {
(3{g_A}^2 +1)m_{\pi}^2 \over (4 \pi f_{\pi})^2} \ln \Bigl( {m_{\pi}^2\over
m_{\pi}^2 + \mu^2} \Bigr) \right] + b_n m_{\pi}^2 \label{chiral_ext}
\end{equation}
The coefficient of the leading non-analytic behavior  $m_{\pi}^2 
ln(m_{\pi}^2)$ is
determined from chiral perturbation theory. The parameter $\mu$ specifies the
scale above which pion loops no longer produce rapid variation. It 
corresponds to the
upper limit of the momentum integration generated physically by the 
inverse size of the
quark core of the nucleon that serves as the source for the pion 
field. As shown in Fig. \ref{fit},
the value $\mu \sim 550$ MeV from ref
\cite{Detmold:2001jb},  which is consistent with the
value required to extrapolate the nucleon magnetic moment and with 
chiral nucleon
models, extrapolates the world's supply of lattice data to the 
experimental values of
$\langle x \rangle_{u-d}$, $\langle x^2 \rangle_{u-d}$, and $\langle x^3
\rangle_{u-d}$.  From this figure, it is also clear that 5\% 
measurements down to
$m_{\pi}^2$ = 0.05 GeV$^2$,
requiring a spatial volume of order (4.3~fm)$^3$,
would provide data for a definitive
lattice calculation.
 From the known scaling of present hybrid Monte Carlo algorithms,
this calculation will require 8 Teraflops-years and thus can be 
carried out on the next
generation of 10-Teraflops computers.
\begin{figure}[t]
\vspace*{+2mm}
$$
\BoxedEPSF{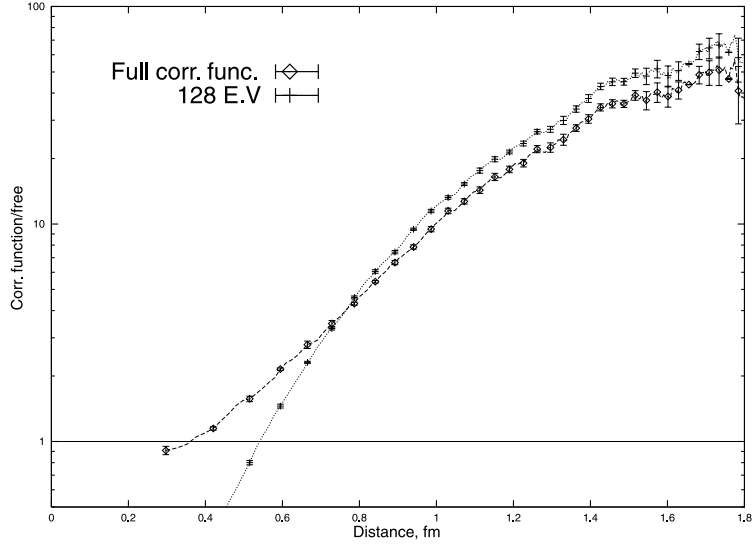 scaled 1100}
$$
\vspace*{-.4cm}
\caption{Graph showing how the propagation of a pion in the QCD 
vacuum (diamonds) is well
approximated by a propagator containing only the lowest 128 eigenmodes}
\label{eigenmode}
\vspace*{-.7cm}
\end{figure}

\section*{FUTURE PROMISE}

With the present tools of lattice field theory and the physical 
understanding of the essential role
of the pion cloud, we now have the framework for definitive 
calculation of hadron structure.
Improved actions will allow a reliable extrapolation to the continuum 
limit.  Chiral fermions now
implement an exact chiral symmetry on the lattice.  Partially 
quenched chiral perturbation
theory provides a systematic framework to measure the relevant 
parameters of effective field
theory for unambiguous chiral extrapolation using optimal 
combinations of valence and sea
quark masses, including corrections for finite volume and fixed 
topology.  As explained above,
computers providing sustained performance of tens of Teraflops will 
enable calculation in the
chiral regime of sufficiently low quark mass and sufficiently large 
volume for reliable chiral
perturbation theory extrapolation. Renormalization coefficients and 
mixing parameters can now
be calculated nonperturbatively. And finally,  eigenmode expansion techniques
\cite{instantons,Neff:2001zr} using the dominance of propagators by 
low eigenmodes as
illustrated in Fig.
\ref{eigenmode} enable the calculation of previously intractable 
disconnected diagrams.

In conclusion, understanding the quark and gluon structure of QCD 
requires not only frontier
facilities for experimental measurements, but also frontier 
facilities for fundamental lattice
calculations.  The time has come to think of dedicated computers in 
the same way as one thinks
of experimental apparatus for the field.  And, as has been so 
effective in experimental physics,
the time has also come for international collaboration in large scale 
computation.

\section*{ACKNOWLEDGMENTS}

It is a pleasure to acknowledge the contributions of
D. Dolgov,  S. Capitani, D. Renner, A. Pochinsky, and  R. Brower, to 
these lattice
calculations,  P. Dreher to the computer cluster infrastructure,
N. Eicker, T. Lippert,  and  K. Schilling in providing the SESAM 
configurations,
R. G. Edwards, and U. M. Heller in providing the SCRI configurations, and
W.~Detmold, W.~Melnitchouk,  and A.~W.~Thomas  in
understanding the chiral extrapolation.


\end{document}